# Observation of wavelength-dependent Brewster angle shift in 3D photonic crystals

Priya and Rajesh V. Nair[*]

Laboratory for Nano-scale Optics and Meta-materials (LaNOM)

Department of Physics, Indian Institute of Technology (IIT), Ropar

Rupnagar, Punjab 140 001 INDIA

[*]Email: rvnair@iitrpr.ac.in

Abstract

The interaction of polarized light with photonic crystals exhibit unique features due to its sub-wavelength nature on the surface and the periodic variation of refractive index in the depth of the crystals. Here, we present a detailed study of polarization anisotropy in light scattering associated with three-dimensional photonic crystals with face centered cubic symmetry over a broad wavelength and angular range. The polarization anisotropy leads to a shift in the conventional Brewster angle defined for a planar interface with certain refractive index. The observed shift in Brewster angle strongly depends on the index contrast and lattice constant. Polarization-dependent stop gap measurements are performed on photonic crystals with different index contrast and lattice constants. These measurements indicate unique stop gap branching at high-symmetry points in the Brillouin zone of the photonic crystals. The inherited stop gap branching is observed for TE polarization whereas that is suppressed for TM polarization as a consequence of Brewster effect. Our results have consequences in the polarized light-scattering from plasmonic structures and dielectric meta-surfaces and are also useful in applications like nano-scale polarization splitters and lasers.

Keywords: Photonic crystals, diffraction and scattering, metamaterials, polarization,

## I.  Introduction

The polarization of light is a peculiar property which is inherited from the vectorial nature of wave equation that governs the light propagation in material media [1]. The polarization of light is defined as the orientation of associated electric (or magnetic) field with respect to the plane of incidence. It is possible to decompose the incident light polarization into two orthogonal components and discern the optical properties based on it. When the electric field is oriented perpendicular to the plane of incidence, it is called the transverse electric (TE) polarization and when the electric field is oriented parallel to the plane of incidence, it is called the transverse magnetic (TM) polarization. The reflectivity and transmittance of light at the planar interface are described on the basis of angle and polarization state of incident light. A set of equations can be derived for the reflection and transmission coefficients by satisfying the boundary conditions for electric and magnetic fields at the interface. This set of equations are specified according to the angle of incidence ($\theta$), polarization state of light, and the difference in refractive index are known as Fresnel equations [2]. This equation suggests that the reflectivity increases with increase in $\theta$ for TE polarized light. However, in the case of TM polarization, the reflectivity decreases first, reaches a minimum value at a certain value of $\theta$, and thereafter it increases. It is interesting to note the zero reflectivity for TM polarization at particular value of $\theta$ known as the Brewster angle ($\theta_B$) which depends on the ratio of refractive index of the two materials. The predictions of Fresnel equations are proved subsequently in many isotropic materials which results in many applications in designing thin film filters, multi-channel filters, polarizers, and Brewster mirror windows [2]. These polarization-dependent optical phenomena are studied and explored for applications considering smooth planar semi-infinite isotropic dielectric materials [3].

Contemporary research interest in the field of nanophotonics has introduced the artificially tailored dielectric materials on the optical wavelength scale called the photonic metamaterials. The Brewster effect was originally interpreted for a homogeneous medium howbeit it now encounters an inhomogeneity at the material interface which entails a modification in the traditional Brewster law for photonic metamaterials [4, 5]. Recent advances in nano-fabrication techniques shows a great possibility in the synthesis of photonic metamaterials with exotic optical properties which are otherwise absent in conventional systems [6]. Photonic crystals are a class of metamaterials wherein the dielectric constant is varied periodically in either two- or three-orthogonal directions [7, 8]. These are generally characterised by direction- and polarization-dependent frequency gaps known as photonic stop gaps wherein certain frequencies of light are forbidden to propagate inside the crystal [9]. The stop gap arises due to the Bragg diffraction of light from the crystal planes in the propagation direction. The careful design of the crystal symmetry with appropriate choice of refractive indices, direction- and polarization-independent frequency gaps are observed called photonic band gap [10]. The consequence of photonic band gap is the rigorous modification of photon density of states in a finite frequency range which affect the fundamental properties of light-matter interactions and have applications in nano-lasers [11], quantum electrodynamics [12], and waveguides [13].

There are other interesting symmetry-dependent optical processes that occur in two- and three-dimensional photonic crystals are the sub-Bragg diffraction [14] and stop gap branching [15-17]. These occur when the incident wave vector passes through certain high-symmetry points in the Brillouin zone of the crystal. The photonic crystals with different symmetry can be synthesized using various approaches like colloidal self-assembly [18], laser-writing [19], and reactive ion-etching [20] methods. Among these, self-assembly approach is more attractive for the synthesis of 3D photonic crystal due to its ease of fabrication and affordability [21]. Such 3D photonic crystals using colloids with sub-micron diameters are always assembled in the face centered cubic (*fcc*) symmetry [22].

The synthesis and optical characterization of 3D self-assembled photonic crystals with *fcc* symmetry is well-documented in literature [21]. It has been discussed that self-assembled photonic crystals do not exhibit a 3D band gap rather possess only photonic stop gaps [21]. Recent interest in the angle- and polarization-resolved stop gaps indicate interesting optical effects like stop gap branching, band repulsion, and inhibition of spontaneous emission [23-29]. It is shown that TE polarized light interacts strongly with the photonic crystal structure and results in stop gap branching whereas it is absent for TM polarized light due to prevailing Brewster effect [29]. A detailed study of polarization-induced stop gap formation has shown a direction- and wavelength-dependent shift in $\theta_B$ at the stop gap wavelength as compared to the conventional definition of $\theta_B$ [17]. Hence there is polarization anisotropy in the light scattering in photonic crystals. However, the influence of dielectric contrast and the role of material refractive index dispersion on polarization anisotropy are yet to be analyzed.

In the present work, we investigate the role of index contrast and the wavelength dispersion of material refractive index on the polarization anisotropy in photonic crystals. We discuss the shift in $\theta_B$ as a function of index contrast. Further, we elucidate the polarization-dependent stop gap branching for photonic crystal in different spectral ranges and for different index contrasts. The present study will amplify the knowledge about strong interaction of polarized light with photonic crystals leading to multiple prospects in photonics and plasmonics. Additionally, the eccentric Brewster effect in sub-wavelength nanophotonic structures has its impact in various applications giving them a new edge with enticing properties, relevant to the polarization of light.

## II. Experimental details

IIA. Sample details

The photonic crystals used in the present work are synthesized using convective self-assembly method [18]. A clean glass substrate is placed vertically in a cuvette containing colloidal suspension of appropriate volume and concentration which is kept inside a temperature controlled oven for 4-5 days. It is well-known that self-assembled photonic crystals grown using sub-micron spheres are always organized into *fcc* geometry as it is the minimum energy configuration [22]. We have chosen photonic crystal samples made of polystyrene (PS) microspheres of diameter 803 ± 20 *nm* (sample *A*) and 287 ± 8 *nm* (sample *B*) to study the role of refractive index dispersion in the origin of polarization anisotropy. Also samples are grown using polymethyl methacrylate (PMMA) microspheres of diameter 286 ± 10 *nm* (sample *C*) to study the impact of index contrast on the polarization anisotropy in photonic crystals having similar lattice constants.

IIB. Measurement details

Scanning electron microscope (SEM) is used to visualize the structure of photonic crystals. The SEM image of our photonic crystal exhibits high quality ordering on the surface and as well as in the depth [16]. The surface shows the hexagonal array of microspheres which depicts the (111) plane of the crystal with *fcc* symmetry [16]. Angle- and polarization-dependent reflectivity measurements are done in the specular reflection geometry using PerkinElmer Lambda 950 spectrophotometer. The sample is mounted in such a way that enables us to access high-symmetry points in the Brillouin zone of crystal with *fcc* symmetry [30]. The light source used is a Tungsten-Halogen lamp. The reflected light is collected using a photomultiplier tube and lead sulfide detector for the visible and near-infrared spectral ranges, respectively. The spot size of beam on the sample is 5 × 5 *mm*. The plane of incidence is perpendicular to the top surface of the crystal. The polarizer (Glan-Thomson; wavelength range: 300-3000 *nm*) is mounted in the incident beam path to select either TE or TM polarization of light.

### III. Results and analysis

IIIA. Calculated stop gap crossing

When the light is incident normally on the samples, the stop gap is formed due to the diffraction of light from the (111) planes of crystal. This corresponds to (111) stop gap in the $\Gamma L$ direction for the crystal with *fcc* symmetry [30]. For higher values of $\theta$ ($\theta > 45°$), the incident wave vector approaches other high-symmetry points in the hexagonal facet of the Brillouin zone of crystal with *fcc* symmetry. In the vicinity of a high-symmetry point, the diffraction conditions can be satisfied for multiple crystal planes in a complex manner leading to interesting optical process like stop gap branching and band crossing [15-17]. The stop gap wavelengths ($\lambda_{hkl}$) from different crystal planes with Miller indices (*hkl*) as a function of $\theta$ can be calculated using the Bragg's law [26]:

$$\lambda_{hkl} = 2n_{eff}d_{hkl}\cos\left[\alpha - \sin^{-1}\left(\frac{1}{n_{eff}}\sin\theta\right)\right], \ldots\ldots\ldots\ldots\ldots\ldots\ldots\ldots\ldots\ldots(1)$$

where $d_{hkl}$ is interplanar spacing, $n_{eff}$ is the effective refractive index, and $\alpha$ is the internal angle between the (*hkl*) and (111) plane.

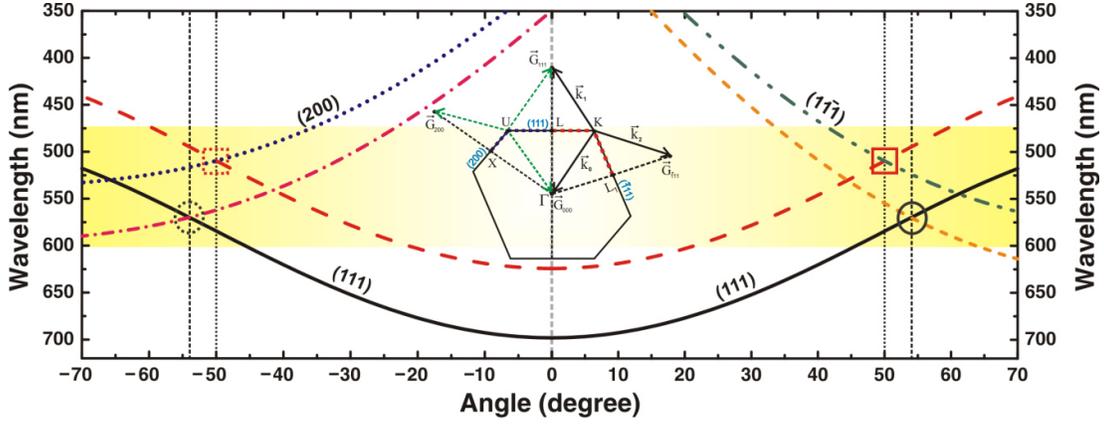

**Figure 1.** The calculated angular dispersion of stop gap wavelengths along the *LK* and *LU* lines with respect to *ΓL* direction. The calculations are shown for samples *B* and *C*. The (111) stop gap for sample *B* (solid line) and sample *C* (dashed line) shifts towards the shorter wavelengths and intercepts either $(\bar{1}11)$ or the (200) stop gap at *K* (solid symbols) or *U* (dotted symbols) high-symmetry point at certain $\theta$. The crossing occurs at the same wavelength for both *K* and *U* points for both samples. The stop gap crossing at *K* (*U*) point for sample *B* occurs at higher wavelength compared to sample *C* due to difference in the index contrast. The inset shows the diffraction conditions that can lead to the origin of multiple stop gaps at the *K* (*U*) point.

Figure 1 shows the calculated stop gap wavelengths at different $\theta$ using eq. (1) for samples *B* and *C*. The samples *B* and *C* are specifically chosen to show the effect of index contrast and thus different value of $n_{eff}$ for photonic crystals having nearly the same lattice constant. The calculations are done for the case when the tip of the incident wave vector spans across the *LK* and *LU* lines in the cross-section of the Brillouin zone (Fig.1 (inset)). The stop gap calculations assume certain value of $n_{eff}$ and $d_{hkl}$ and their estimation will be discussed in later section IIIC. The (111) stop gap for sample *B* (solid) and sample *C* (dashed) is shifted towards shorter wavelength side. The $(\bar{1}11)$ stop gap for sample *B* (short dashed) and for sample *C* (dash dot-dot) is shifted towards the longer wavelength side. The (200) stop gap for sample *B* (dash-dot) and sample *C* (dotted) also shift towards longer wavelength side with increase in $\theta$. The (111) stop gap intercepts $(\bar{1}11)$ and (200) stop gaps at *K* (solid symbols) and *U* (dotted symbols) high-symmetry points, respectively for the same value of $\theta$ for a given photonic crystal. This is due to the fact that the length of *LK* and *LU* lines are equal on the hexagonal facet of Brillouin zone of the crystal with *fcc* symmetry [23]. The stop gap crossing at both *K* and *U* points occur at different $\theta$ for sample *B* (circles) and *C* (squares) due to the difference in their index contrast. The stop gap crossing for sample *C* occurs at much earlier $\theta$ value which signifies the role of $n_{eff}$ in the crossing $\theta$ value.

The stop gaps crossing at certain $\theta$ clearly suggests the presence of multiple diffraction resonances due to (111) and $(\bar{1}11)$ or (111) and (200) planes depending on the *K* or *U* high-symmetry point. The formation of multiple diffraction resonances can be explained through diffraction conditions at the *K* (*U*) point as shown in Fig.1 (inset). When the wave vector is incident along *ΓK* direction, diffraction conditions are satisfied for (111) and ($\bar{1}$11) planes simultaneously with conditions $\vec{k}_0 + \vec{G}_{111} = \vec{k}_1$ and $\vec{k}_0 + \vec{G}_{\bar{1}11} = \vec{k}_2$, where $\vec{k}_1$ and $\vec{k}_2$ are the diffracted wave vectors from the (111) and ($\bar{1}$11) planes with reciprocal vectors $\vec{G}_{111}$ and $\vec{G}_{\bar{1}11}$, respectively. In a similar way, diffraction conditions can also emanate at the *U* point. This process is known as multiple Bragg diffraction which is accompanied with the branching of stop gaps in the reflectivity or transmission measurements [27]. It is also noticed that the stop gap branching eminently depends on the polarization states of light due to the unique vectorial nature of light as discussed below.

IIIB. Optical reflectivity measurements

We have performed an extensive set of angle- and polarization-resolved optical reflectivity measurements on photonic crystals. At near-normal incidence ($\theta = 10°$), which corresponds to *L* point, the reflectivity spectra show (111) stop gap for sample *A* and *B* centered at 1770 ± 2 *nm* and 667 ± 9 *nm*, respectively and that for sample *C* occur at 593 ± 7 *nm*. The measured stop gap widths is in complete agreement with calculated photonic band structure at near-normal incidence [30]. The reflectivity spectra for the samples clearly show the Fabry-Perot (F-P) fringes in the long-wavelength limit. The thickness estimated from these F-P fringes is 13 ± 0.37 *μm* (~20 layers) for sample *A*, 8 ± 0.47 *μm* (~32 layers) for sample *B*, and 6.4 ± 0.92 *μm* (~27 layers) for sample *C*.

To elucidate the stop gap formation at higher $\theta$ in our measurements, the crystal is oriented in such a way to scan the stop gap along the *LK* or *LU* line in the Brillouin zone. It is seen in section IIIA that at higher values of $\theta$, the incident wave vector passes through the high-symmetry points which results in the excitation of new diffraction resonances from crystal planes other than the (111) plane. There is a considerable debate in literature regarding the assignment of planes responsible for the origin of new stop gaps at high-symmetry points [15, 16, 26, 30].

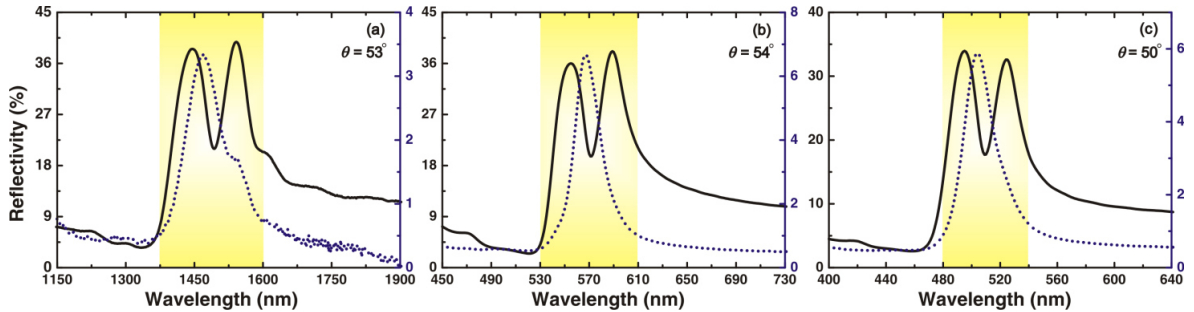

**Figure 2.** The reflectivity spectra measured using TE (solid) and TM (dotted) polarized light at certain $\theta$ value which relates the *K* or *U* high-symmetry point for the respective photonic crystal samples. Fig. 2(a), 2(b), and 2(c) show the reflectivity spectra at $\theta = 53°$, 54°, and 50° for sample *A*, *B*, and *C*, respectively. The TE polarized reflectivity spectra show a secondary peak in addition to the (111) stop gap for all the samples. In contrast, the TM polarized reflectivity spectra show only (111) stop gap with a very low reflectivity value.

Figure 2 depicts the TE (solid) and TM (dotted) polarized reflectivity spectra at selected $\theta$ values corresponding to *K* (*U*) point for all the samples. The reflectivity spectra are shown at $\theta = 53°$ for sample *A*, at $\theta = 54°$ for sample *B*, and at $\theta = 50°$ for sample *C* in Fig.2 (a-c), respectively. The stop gap branching is evident for all the crystals for TE polarized light. However, only (111) stop gap is present for TM polarized light that appears mid-way between the two TE polarized stop gaps. The difference in peak reflectivity values is due to the different thickness of the samples in addition to the difference in index contrast. The (111) and the new (*hkl*) stop gap for TE polarization shows nearly equal reflectivity (~40% for sample *A*, ~38% for sample *B*, and ~34% for sample *C*) with a separation of 96 *nm* for sample *A*, 35 *nm* for sample *B*, and 30 *nm* for sample *C*. The separation between peaks are greater than their individual line-width (~60 *nm* for sample *A*, ~24 *nm* for sample *B*, and ~20 *nm* for sample *C*) suggests the signature of strong coupling between the two TE polarized stop gaps. The branching occurs for sample *A* and *B* at nearly the same $\theta$ value due to slight difference in the $n_{eff}$ value. However, sample *C* shows the stop gap branching at a much less $\theta$ value due to the reduced value of $n_{eff}$. In contrast, the TM polarized reflectivity spectra show only (111) stop gap with extremely low reflectivity values for all the samples (3.5% for sample *A*, 7% for sample *B*, and 5.7% for sample *C*). This indicates the weak interaction between TM polarized light and photonic crystals. The reflectivity values of (111) stop gap for TM polarization is reduced significantly due to the Brewster effect. The Brewster effect also accounts for the suppression of energy exchange between planes in the depth of the crystal and results in the absence of stop gap branching. Therefore, the polarization-induced stop gap branching is a generic behaviour exhibited by photonic crystals irrespective of their lattice constant and index contrast.

IIIC. Estimation of $n_{eff}$ and $d_{hkl}$ from measurements

The calculation of stop gap wavelengths using eq. (1) at different $\theta$ is performed using certain values of $n_{eff}$ and *D*. The precise extraction of $n_{eff}$ in a photonic crystal structure is quite complicated and many models had explored [31]. However, we use a unique way to estimate $n_{eff}$ value from the measured reflectivity spectra [17, 31]. This way of estimating $n_{eff}$ is quite useful in explaining many optical effects associated with self-assembled photonic crystals.

Assume that the stop gap branching discussed in section IIIB for TE polarization occurs for the wave vector incident along the $\Gamma K$ direction (see Fig.1 (inset)) in the Brillouin zone. The internal angle between $\Gamma L$ and $\Gamma K$ directions is 35.5° and applying Snell's law to this geometry we can estimate $n_{eff} = \sqrt{3} \sin \theta_K$. The value of $\theta_K$ corresponds to the $\theta$ value at which the stop gap branching occurs in the measured reflectivity spectra for TE polarized light. The estimated value of $n_{eff}$ for sample *A* is 1.38, for sample *B* is 1.40, and for sample *C* is 1.33. The sample *B* has the highest $n_{eff}$ and therefore, it also has the highest index contrast among our samples. The reflectivity spectra at $\theta = \theta_K$ show a clear trough between the two peaks. This suggests that the light is transmitted through the sample and the crystal structure as a whole behaves like a homogenous material at that particular wavelength ($\lambda_K$). Therefore, we can use the free photon dispersion relation which can be written as,

$$\omega = \frac{c|\overrightarrow{\Gamma K}|}{n_{eff}} \quad \ldots\ldots\ldots\ldots\ldots\ldots\ldots\ldots\ldots\ldots\ldots (2),$$

where $|\overrightarrow{\Gamma K}| = \sqrt{\frac{3}{2}}\frac{\pi}{d_{111}}$ is the length of the incident wave vector at $\theta_K$, $d_{111} = 0.816D$, and *c* is the speed of light. In eq. (2), substituting $\omega$ in terms of $\lambda_K$ as $\omega = \frac{2\pi c}{\lambda_K}$, gives the value of *D* as

$$D = \frac{3\lambda_K}{4n_{eff}} \quad \ldots\ldots\ldots\ldots\ldots\ldots\ldots\ldots\ldots\ldots\ldots (3),$$

The estimated value of *D* for sample *A* is 809 *nm*, for sample *B* is 305 *nm*, and for sample *C* is 288 *nm*. In order to check the reliability of our estimation, we have calculated stop gap wavelength using Bragg's law at near-normal incidence for sample *B*. The calculated stop gap wavelength is 696 *nm* which is in close agreement with measured value. Hence, using our estimated value of $n_{eff}$ and *D*, the angular dispersion of stop gap is calculated and compared with the measurements; as discussed below.

IIID. Comparison between the measured and calculated stop gaps

Figure 3 shows the measured (symbols) and calculated (lines) stop gap wavelengths using TE (left panel) and TM (right panel) polarized light as a function of $\theta$. The calculations are done for (111) [solid], ($\bar{1}$11) [dashed], and (200) [dotted] planes which are the only relevant planes that can intersect at the *K* or *U* point. The measured (111) stop gap (squares) wavelengths are in close agreement with the calculations, till the appearance of new stop gaps (circles) for TE polarization. The measured (111) stop gap appears to be shifted away from the calculated curve and avoidably crosses the new stop gap at calculated crossing $\theta$ value. Beyond crossing, the new stop gap is in good agreement with the ($\bar{1}$11) stop gap. In contrast, the stop gap for TM polarization follows the calculated (111) stop gap wavelengths at all $\theta$ values. However, the new stop gap (circles) which appears well above the crossing $\theta$ value is in good agreement with calculated ($\bar{1}$11) stop gaps. The ($\bar{1}$11) stop gap for TM polarization emerges at $\theta$ that relates with $\theta_B$ for a given photonic crystal and it is only due to the Brewster effect that the ($\bar{1}$11) stop gap is absent for any $\theta < \theta_B$ for TM polarization. The polarization-dependent stop gap branching is illustrated here for photonic crystals having different index contrast but with different lattice constant.

Figure 3 confirms the good agreement between the measured new stop gaps and the calculated ($\bar{1}11$) stop gaps irrespective of photonic crystal samples for both TE and TM polarization. This strong agreement guarantees that the new reflectivity peak is ($\bar{1}11$) stop gap and the incident wave vector is shifted along a line connecting the *L* and *K* points in the Brillouin zone. It is worth mentioning that the $n_{eff}$ and *D* used in the calculations are estimated from measured reflectivity spectra assuming that the light is incident along the *ΓK* direction. This atypical polarization-induced stop gap branching at different spectral ranges and for crystal with different index contrast is addressed here for the first time.

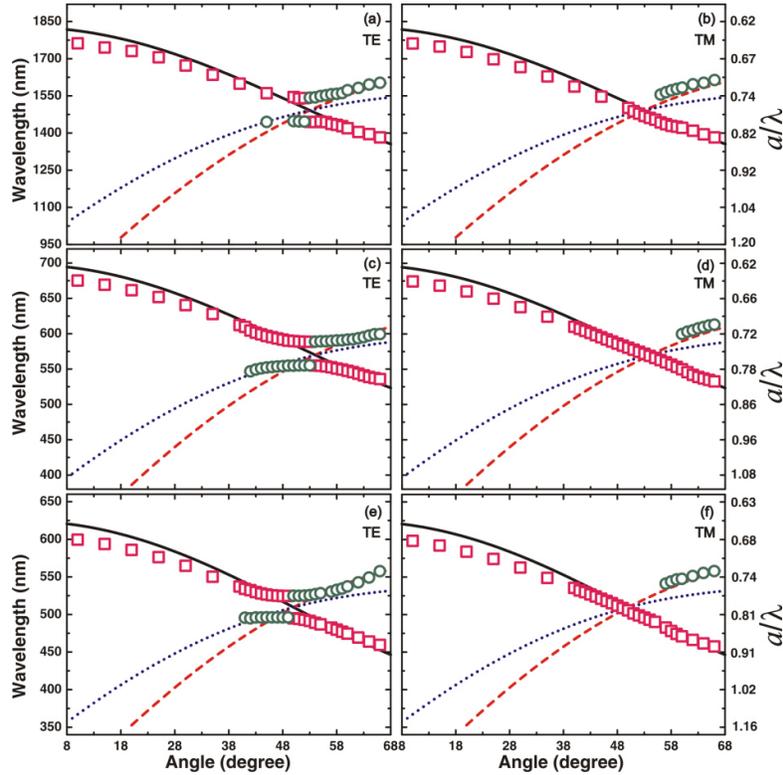

**Figure 3.** The measured (symbols) and calculated (lines) stop gap wavelengths at different *θ* for TE (left panel) and TM (right panel) polarized light. The planes used in the calculations are (111) [solid line], ($\bar{1}11$) [dashed line], and (200) [dotted line]. The stop gap branching is observed for a range of *θ* for all the samples for TE polarization. The (111) stop gap for TE polarization appears to be moving away from the calculated curve with the emergence of new stop gap (circles) for all samples. This band repulsion is observed due to the avoided crossing of stop gaps at different values of *θ*. In contrast, no such avoided crossing is observed for TM polarization for all the samples. The new stop gap emerges beyond the crossing angle for $\theta > \theta_B$ in the case of TM polarization. The new stop gap appeared beyond crossing angle is in complete agreement with calculated ($\bar{1}11$) stop gaps for both TE and TM polarization for all samples.

The stop gap branching is observed for a broad angular and wavelength range for TE polarization using samples with different index contrast as seen in Fig.3. Apparently, the measured (111) stop gap is repelled away from the calculated curve near the crossing *θ* value. As the (111) stop gap approaches the exact crossing angle ($\theta_K$), the band repulsion occurs due to the presence of ($\bar{1}11$) stop gap. The

(111) and ($\bar{1}$11) stop gaps are diffraction resonance modes which are trying to appear at the same wavelength and therefore exhibit an avoided crossing behaviour. However, for TM polarization, stop gap branching is not at all observed till $\theta = \theta_B$ and the ($\bar{1}$11) stop gap appears for $\theta > \theta_B$ for all the samples. This suggests that the stop gap branching is absent due to the prevailing Brewster effect for TM polarized light. We have clearly observed the origin of ($\bar{1}$11) stop gap for TM polarization at $\theta = \theta_B$ (i.e. far away from the crossing regime) for all the samples. It is interesting to note the appearance of ($\bar{1}$11) stop gap at different $\theta_B$ values for sample *A*, *B*, and *C* which is due to the difference in the value of $n_{eff}$. Since the estimated $n_{eff}$ value for sample *A* and *B* is nearly equal results in nearly the same $\theta_K$ value whereas the $\theta_K$ value for sample *C* is much lower due to the reduced value of $n_{eff}$.

### IV.    Measured and calculated polarization anisotropy

It is seen that the polarization-dependent angular dispersion of stop gaps for all the samples is differentiated on the basis of $n_{eff}$ and difference in their index contrast. Such unique polarized light interaction is not only due to 3D periodicity within the bulk of the crystal but also inherited to sub-wavelength nature of top crystal surface. To quantitatively explain the anomalies with regard to the TE and TM polarized stop gaps, we estimate a factor called polarization anisotropy ($P_a$) [29]. It is defined as the ratio of reflectivity of TM to TE polarized light at a certain wavelength for a given $\theta$ value. Fig.4 (a-c) show the measured $P_a$ values at different $\theta$ for off-resonance wavelengths (squares) which corresponds to $D/\lambda = 0.39$ and for on-resonance wavelengths that corresponds to (111) stop gap (circles) for samples *A*, *B*, and *C*, respectively. Fig.4 (a-c) also show the calculated $P_a$ values (dotted line) using Fresnel equations assuming the sample as a thin film with respective $n_{eff}$ value. This is a good approximation in the long-wavelength limit of our photonic crystals. Fig.4 (a-c) inset show the $P_a$ values for the on-resonance wavelengths corresponding to ($\bar{1}$11) stop gap (diamonds) for sample *A*, *B*, and *C*, respectively.

It is to be noted that the measured and calculated $P_a$ values is the same at $\theta \sim 0°$ due to the identical nature of TE and TM polarization. The $P_a$ value decreases first, reaches a minimum at certain $\theta$, and then again increases with increase in $\theta$ for all the samples. The calculated and the measured off-resonance $P_a$ value show the minimum at the same $\theta$ value. This is because the crystal structure responds to the incident light at off-resonance wavelength (long-wavelength limit) as if it is a homogeneous medium. This minimum $P_a$ value corresponds to $\theta_B$ of the photonic crystal which is related to the value of $n_{eff}$ and for $\theta > \theta_B$, the $P_a$ value increases with increase in $\theta$ as expected from Fresnel equation. The calculated and measured off-resonance $\theta_B$ clearly correspond to the same value of $n_{eff}$ which further validates our estimation of $n_{eff}$.

It is astonishing to observe the shift in $\theta_B$ for on-resonance wavelength as compared to off-resonance wavelength and the calculated $\theta_B$ value. This clearly suggests that the on-resonance $P_a$ minimum is shifted to the higher side due to the complex nature of $n_{eff}$ at the on-resonance wavelength and light scattering at the photonic crystal surface. This also shows that there is a competition set between the Brewster effect and the Bragg diffraction due to which the Brewster angle is shifted to higher $\theta$ values for on-resonance wavelength. The shift in the value of $\theta_B$ states the wavelength-dependent Brewster effect in photonic crystals due to the strong dispersion of $n_{eff}$. The definition of $n_{eff}$ at the stop gap wavelength is a subtle issue which further complicates the interpretation of shift in $\theta_B$ in addition to the role played by the fabrication-induced intrinsic disorder. Hence advanced theoretical formulation and simulations are required by taking care of the sub-wavelength nature of surface which is beyond the scope of the present work.

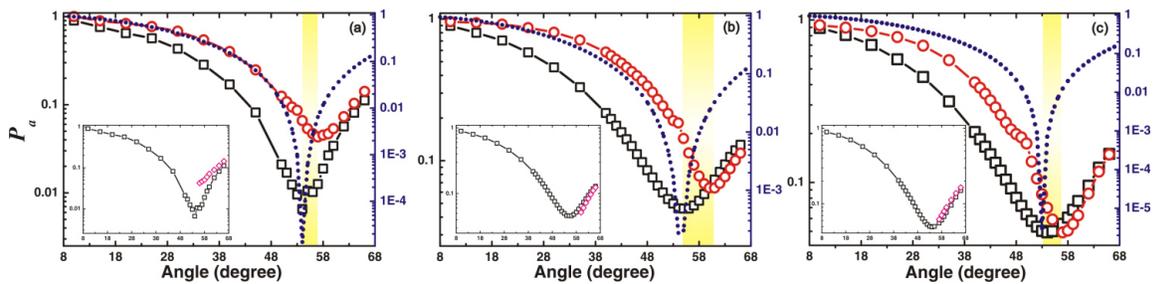

**Figure 4.** The polarization anisotropy factor ($P_a$) from the measured reflectivity spectra is shown for (a) sample $A$, (b) sample $B$, and (c) sample $C$. The $P_a$ values are shown for off-resonance wavelength (squares) for which $D/\lambda = 0.39$ and for on-resonance wavelength (circles) corresponds to (111) stop gap wavelength for all the samples. The calculated $P_a$ values (dotted line) assuming the crystal as a material of certain $n_{eff}$ using Fresnel equations is also shown for the respective samples. The $P_a$ values show a decrease with increase in $\theta$, achieves a minimum at $\theta_B$, and then increases for $\theta > \theta_B$. The calculated and the off-resonance $P_a$ values show the minimum at the same $\theta$ value. However, the on-resonance minimum $P_a$ value is shifted to the higher $\theta$ value for all the samples. The shift in the on-resonance $P_a$ minimum is also confirmed for ($\bar{1}11$) stop gap for all the samples as shown in the inset.

The on-resonance $P_a$ value for the ($\bar{1}11$) stop gap also shows a shift in $\theta_B$ for all the samples as compared to the calculated and off-resonance value as seen in Fig.4 (a-c) inset. However, the observed shift in $\theta_B$ is not constant rather it depends on the index contrast and lattice constants of photonic crystals. The $P_a$ values depicted in Fig.4 (inset) clearly show the absence of ($\bar{1}11$) peak for $\theta < \theta_B$. This is due to the fact that the process of energy exchange between planes within the crystal is inhibited by the Brewster effect and therefore the ($\bar{1}11$) plane is not able to diffract leading to the absence of ($\bar{1}11$) stop gap for $\theta < \theta_B$.

The measured shift in the on-resonance $P_a$ minimum is different for all the samples which are related to the difference in the value of $n_{eff}$. The measured shift in $\theta_B$ for sample $B$ is 6° which is in accordance with the theoretical calculations done for similar crystals with an ideal *fcc* lattice symmetry [29]. The shift in $\theta_B$ observed for samples *A* and *C* is 3°. The shift in $\theta_B$ varies with $n_{eff}$ in the diffraction regime due to its wavelength dispersion for sample *A* and *B* and change in index contrast for sample *C*. The sample *B* is having the largest index contrast as it owes the highest $n_{eff}$ value among our samples which ensues a large shift in $\theta_B$. The sample *C* shows smallest shift in $\theta_B$ as it posses least $n_{eff}$ value among the samples and therefore has the lowest index contrast. This clearly shows that the shift in $\theta_B$ strongly depends on the index contrast. This further supports the role played by the index contrast in photonic crystal; larger the contrast stronger will be the light-matter interaction.

## V. Conclusions and perspective

We have shown the influence of refractive index contrast and lattice constants in the origin of polarization anisotropy in 3D photonic crystals with *fcc* symmetry. The wavelength-dependent polarization anisotropy is studied using the angle-and polarization-resolved stop gaps along the *Γ-L-K-$L_1$* path in the Brillouin zone. The TE polarized light shows stop gap branching in the vicinity of *K* point which is due to multiple Bragg diffraction from the (111) and ($\bar{1}11$) planes. The stop gap branching can also be explained as an inflow of energy between the (111) and ($\bar{1}11$) planes in the depth of the crystal. In contrast, we have observed that the stop gap branching is reluctant to TM polarized light in crystals with *fcc* symmetry. This is due to the prevailing Brewster effect that inhibits the energy exchange within the crystal. Beyond $\theta_B$, a new stop gap is formed which further confirms the role of energy exchange in the formation of secondary stop gap in crystal with *fcc* symmetry. At any rate, the new stop gap formed is in good agreement with the calculated ($\bar{1}11$) stop gap for TE and TM polarizations and the incident wave vector is shifted towards the *K* point in the Brillouin zone.

It is observed that stop gap repulsion is occurred leading to the avoided crossing in the vicinity of *K* point for TE polarization. Conversely, no such stop gap repulsion is observed for TM polarization and hence the avoided crossing of the stop gaps is absent. The (111) stop gap for TM polarization is in complete agreement with calculated stop gaps for all *θ*. However, the stop gap branching originates at much higher *θ* value due to the presence of ($\bar{1}11$) stop gap for TM polarization.

The avoided crossing occurs at different values of $\theta_K$ for all the photonic crystal samples considered in our work. We have observed a shift of 4° in $\theta_K$ for sample *C* compared to sample *B* which is ascribable to the difference in index contrast even though the samples have nearly equal value of lattice constants. A minor shift in the crossing angle observed for sample *A* as compared to that for sample *B* even though both are made using same material. By virtue of dispersion properties of PS polymer, the stop gap wavelengths occurring in different wavelength regimes experience a different $n_{eff}$ for the photonic crystal structure. The PS photonic crystal with larger lattice constant show stop gap in the near-infrared wavelength range where as one with smaller lattice constant in the visible range. Consequently, the crossing is achieved earlier for the medium that proffers lower $n_{eff}$ value.

The measured $P_a$ factor validates the shift in $\theta_B$ for the on-resonance wavelengths as compared to the off-resonance wavelengths for all the samples in our work. However, the shift observed for different samples in our samples is not the same which is due to the difference in the index contrast of crystals. The largest shift in $\theta_B$ (~6°) is observed for sample $B$ which has the highest index contrast. This notion of change in $\theta_B$ strongly depends on the wavelength region of interest which depicts the value of $n_{eff}$. This anomalous shift in $\theta_B$ supports that the shift is influenced by the index contrast and hence the photonic strength of the photonic metamaterials in general.

Our results establish a strong interaction of polarized light with photonic crystals that can be tuned with regard to the refractive index contrast and lattice constant of the crystal. Moreover, the strong wavelength-dependent character of $\theta_B$ in photonic crystal will help in designing them for various applications based on polarization. Furthermore, the study advances the modification of Brewster effect in photonic metamaterials like photonic crystals which is also instructive for other sub-wavelength dielectric photonic structures.

## Acknowledgements

The authors acknowledge the use of UV-Vis-NIR spectrophotometer (Central Facility) at IIT Ropar. The financial support from SERB, DST, Govt. of India and IIT Ropar for internal research support is gratefully acknowledged. Priya acknowledges IIT Ropar for PhD fellowship.

## References

1. Jackson J D (ed) 1998 *Classical Electrodynamics* (NY: John Wiley & Sons)
2. Hecht E (ed) 2002 *Optics* (MA: Addison-Wesley)
3. Shen Y, Hsu C W, Yeng Y X, Joannopoulos J D and Soljacic M 2016 Broadband angular selectivity of light at the nanoscale: Progress, applications and outlook *Appl. Phys. Rev.* **3** 011103
4. Paniagua-Dominguez R, Yu Y F, Miroschnichenko A E, Krivitsky L A, Fu Y H, Valuckas V, Gonzaga L, Toh Y T, Kay A Y S, Luk'yanchuk B and Kuznetsov A I 2016 Generalized Brewster effect in dielectric metasurfaces *Nat. Commun.* **7** 10362
5. Lee K J and Kim K 2011 Universal shift of the Brewster angle and disorder-enhanced delocalization of *p* waves in stratified random media *Opt. Express* **19** 20817
6. Soukoulis C M and Wegener M 2011 Past achievements and future challenges in the development of three-dimensional photonic metamaterials *Nat. Photon.* **5** 523-530
7. Yablonovitch E 1987 Inhibited Spontaneous Emission in Solid-State Physics and Electronics *Phys. Rev. Lett.* **58** 2059

8.  John S 1987 Strong localization of photons in certain disordered dielectric superlattices *Phys. Rev. Lett.* **58** 2486
9.  Lopez C 2003 Material aspects of photonic crystals *Adv. Mater.* **15** 1679
10. Huisman S R, Nair R V, Woldering L A, Leistikow M D, Mosk A P and Vos W L 2011 Signature of a three-dimensional photonic band gap observed on silicon inverse woodpile photonic crystals *Phys. Rev. B* **83** 205313
11. Nair R V, Tiwari A K, Mujumdar S and Jagatap B N 2012 Photonic-band-edge-induced lasing in self-assembled dye-activated photonic crystals *Phys. Rev. A* **85** 023844
12. Lodahl P, Floris van Driel A, Nikolaev I S, Irman A, Overgaag K, Vanmaekelbergh D and Vos W L 2004 Controlling the dynamics of spontaneous emission from quantum dots by photonic crystals *Nature* **430** 654-657
13. Foresi J S, Villeneuve P R, Ferrera J, Thoen E R, Steinmeyer G, Fan S, Joannopoulos J D, Kimerling L C, Smith H I and Ippen E P 1997 Photonic-bandgap microcavities in optical waveguides *Nature* **390** 143-145
14. Huisman S R, Nair R V, Hartsuiker A, Woldering L A, Mosk A P and Vos W L 2012 Observation of sub-Bragg diffraction of waves in crystals *Phys. Rev. Lett.* **108** 083901
15. van Driel H M and Vos W L 2000 Multiple Bragg wave coupling in photonic band-gap crystals *Phys. Rev. B* **62** 9872
16. Nair R V and Jagatap B N 2012 Bragg wave coupling in self-assembled opal photonic crystals *Phys. Rev. A* **85** 013829
17. Priya and Nair R V 2016 Polarization-selective branching of stop gaps in three-dimensional photonic crystals *Phys. Rev. A* **93** 063850
18. Jiang P, Bertone J F, Hwang K S and Colvin V L 1999 Single-crystal colloidal multilayers of controlled thickness *Chem. Mater.* **11** 2132-2140
19. Deubel M, von Freymann G, Wegener M, Pereira S, Busch K and Soukoulis C M 2004 Direct laser writing of three-dimensional photonic-crystal templates for telecommunications *Nature Mater.* **3** 444-447
20. Woldering L A, Tjerkstra R W, Jansen H V, Setija I D and Vos W L 2008 Periodic arrays of deep nanopores made in silicon with reactive ion etching and deep UV lithography *Nanotechnology* **19** 145304
21. Galisteo-Lopez J F, Ibisate M, Sapienza R, Froufe-Perez L S, Blanco A and Lopez C 2011 Self-assembled photonic structures *Adv. Mater.* **23** 30-69
22. Woodcock L V 1997 Entropy difference between the face-centered cubic and hexagonal close-packed structures *Nature* **385** 141-143
23. Andreani L C, Balestreri A, Galisteo-López J F, Galli M, Patrini M, Descrovi E, Chiodoni A, Giorgis F, Pallavidino L and Geobaldo F 2008 Optical response with threefold symmetry axis on oriented microdomains of opal photonic crystals *Phys. Rev. B* **78** 205304
24. Balestreri A, Andreani L C and Agio M 2006 Optical properties and diffraction effects in opal photonic crystals *Phys. Rev. E* **74** 036603
25. Galisteo-Lopez J F, Lopez-Tejeira F, Rubio S, Lopez C and Sanchez-Dehesa J 2003 Experimental evidence of polarization dependence in the optical response of opal-based photonic crystals *Appl. Phys. Lett.* **82** 4068
26. Baryshev A V, Khanikaev A B, Uchida H, Inoue M and Limonov M F 2006 Interaction of polarized light with three-dimensional opal-based photonic crystals *Phys. Rev. B* **73** 033103

27. Baryshev A V, Khanikaev A B, Fujikawa R, Uchida H and Inoue M 2007 Polarized light coupling to thin silica-air opal films grown by vertical deposition *Phys. Rev. B* **76** 014305
28. Romanov S G 2007 Anisotropy of Light Propagation in Thin Opal Films *Phys. Solid State* **49** 536
29. Romanov S G, Peschel U, Bardosova M, Essig S and Busch K 2010 Suppression of the critical angle of diffraction in thin-film colloidal photonic crystals *Phys. Rev. B* **82** 115403
30. Galisteo-Lopez J F, Palacios-Lidon E, Castillo-Martinez E and Lopez C 2003 Optical study of the pseudogap in thickness and orientation controlled artificial opals *Phys. Rev. B* **68** 115109
31. Avoine A, Hong P N, Frederich H, Frigerio J-M, Coolen L, Schwob C, Nga P T, Gallas B and Maitre A 2012 Measurement and modelization of silica opal reflection properties: Optical determination of the silica index *Phys. Rev. B* **86** 165432